\documentclass[sn-apa]{sn-jnl}

\usepackage{url}
\usepackage{amsmath}
\usepackage{eurosym}
\usepackage{balance}
\usepackage{tabularx}
\usepackage{booktabs}
\usepackage{placeins}

\newcommand{\change}[1]{\textcolor{black}{#1}}

\jyear{2023}
\begin{document}

\title[Recent Trends in Cross-Border Data Access by Law Enforcement Agencies]{Recent Trends in Cross-Border Data Access by Law Enforcement Agencies}

\author*[1]{\fnm{Jukka} \sur{Ruohonen}}\email{juanruo@utu.fi}
\affil*[1]{\orgdiv{Faculty of Technology}, \orgname{University of Turku},
\orgaddress{\street{Turun yliopisto}, \city{Turku}, \postcode{FI-20014}, \country{Finland}}}

\abstract{Access to online data has long been important for law enforcement
  agencies in their collection of electronic evidence and investigation of
  crimes. These activities have also long involved cross-border investigations
  and international cooperation between agencies and jurisdictions. However,
  technological advances such as cloud computing have complicated the
  investigations and cooperation arrangements. Therefore, several new laws have
  been passed and proposed both in the United States and the European Union for
  facilitating cross-border crime investigations in the context of cloud
  computing. These new laws and proposals have also brought many new legal
  challenges and controversies regarding extraterritoriality, data protection,
  privacy, and surveillance. With these challenges in mind and with a focus on
  Europe, this paper reviews the recent trends and policy initiatives for
  cross-border data access by law enforcement agencies.}


\maketitle


\section{Introduction}

Law enforcement agencies (LEAs) \change{have a} need to access online data for
their evidence gathering \change{during} crime investigations. Not only does
cyber crime involve online investigations, but also conventional, non-cyber
crime \change{today} often includes online activities and thus requires
gathering electronic evidence. \change{In fact, a 2018 report from the European
  Commission stated that as much as 85\% of all criminal investigations required
  some form of electronic evidence gathering.}\footnote{~SWD/2018/118 final.}
\change{At the same time, technological developments have hindered crime
  investigations by LEAs. In recent years particularly the rise of cloud
  computing has been a major concern. While the United States (US) has already
  passed a law for addressing the issues with cloud computing, also the European
  Union~(EU) has recently agreed on a similar law. This paper provides an
  overview of the recent developments, focusing on cross-border data access from
  a European perspective.}

\change{The paper is a literature review. This approach is justifiable because
  there is already an extensive literature base on the topic. Because the topic
  also spans multiple disciplines, from computer science to political science
  and law, the approach taken is specifically an integrative review; the goal is
  to assess and synthesize the existing literature in a way that allows new
  perspectives to emerge~\citep{Snyder19}. In particular, therefore, the goal is
  not to engage in detailed legal analysis; as will be shown, there are already
  plenty of existing works that have done good jobs at this front. It is
  important to synthesize these works in order to better understand the academic
  record on the topic.}

\change{It is possible to further motivate the paper and its contribution from
  multiple perspectives. Not only is cloud computing a challenge for cyber crime
  investigations and criminal law, but similar challenges are present also in
  terms of economics and industrial policy, cyber security, digital rights, and
  data protection. Increasingly, furthermore, cloud computing is an important
  tenet in the geopolitical rivalry between
  countries~\citep{Brancolini23}. Thus, the so-called ``loss of location'' often
  discussed in the criminal law literature apply also to international relations
  between states and their sovereignties---some authors have even talked about a
  new post-Westphalian system~\citep{Guarda15}. In Europe these geopolitical
  aspects have correlated with aspirations to build sovereign European clouds in
  order to decrease dependencies from other major powers. However, the loss of
  location and its relation to state sovereignty has posed major challenges to
  jurisprudence} because jurisdictions are still largely based on territories of
sovereign states and their supranational unions and alliances. \change{To
  address the challenges, as will be elaborated, both the US and the EU have
  resorted to extraterritoriality, allowing LEAs in their jurisdictions to
  access data stored in foreign locations. A further motivating point can be
  made with respect to technological surveillance.}


Electronic evidence collected by LEAs should not be equated to surveillance
conducted by intelligence agencies\change{---the latter deals with national
  security, whereas the former is bound by criminal law. Despite this
  fundamental division, both agencies can be criticized from a perspective of
  accountability and transparency. As} will be elaborated, the current practices
and legal arrangements for cross-border data access by LEAs are far from
perfect; it remains debatable how accountable and transparent the practices and
arrangements are. Europe is not an exception in this regard. \change{In this
  regard, it should be recalled that} accountability and transparency are
important foundational concepts of liberal democracies. The latter is a
requirement for the former; among other things and ever since the Age of
Enlightenment, transparency has been seen to provide the means by which people
can evaluate the use of power by governments and their administrations
\citep{Molnar20, Moses17}. It is also necessary for monitoring the effectiveness
and efficiency of public policies and public administration. Alas, surveillance
is largely excluded; accountability and non-transparency are often justified by
merely pointing out that the rule of law is followed and secrecy is
necessary~\citep{Setty15}. For instance, in terms of transparency, only a few
countries release some vague statistical reports on their intelligence agencies'
use of lawful interception techniques based on signals intelligence and deep
packet inspection. Even then, such reports hardly qualify as evidence of
accountability~\citep{Molnar20}. The secrecy is often even higher with other
information gathering techniques. The situation is better on the side of law
enforcement agencies whose accesses to user data usually need to go through
judicial processes. In addition, several accountability measures for LEAs and
their data sharing have been established in the European Union~\citep{Aden18}.

Second: surveillance is an important concept from a perspective of the
widespread private sector monitoring of online activities, which, as was shown
already by the Snowden revelations, is closely tied to state
surveillance~(\citealt[][pp.~77--79]{Naef23}; \citealt{Kamp21}). Recently,
authors such as \citet{Zuboff19} have even talked about specific surveillance
capitalism that provides new means of production based on the collection of
personal data and the use of algorithms to transform the data collected into
profits. Much of the European data protection efforts have centered on shielding
people from such surveillance practices of companies. \change{Yet}, every coin
has two sides; at the same time, state surveillance has increased also in
Europe. Due to terrorism, cyber crime, and other factors, there has also been a
political push in Europe to increase the powers of LEAs and decrease the data
protection scrutiny of these~\citep{Treiber22}.

To this end, many technology companies have started to adopt pro-active
accountability and transparency practices by releasing so-called transparency
reports on data requests made by LEAs and, in some cases, intelligence
agencies. Such practices align with the more traditional ideals of corporate
social responsibility. In other words, voluntary transparency reporting of many
large companies covers also such areas as finance, taxes, sustainability and
environmental practices, socially responsible investments, labor rights, and
work conditions.

Third: surveillance is revealing because its secrecy prevents from understanding
what type of data is actually collected. A similar concern applies to evidence
collection by LEAs. As asked by \citet{Carrera15} as well as
\citet{Westmoreland15}, what exactly qualifies as ``evidence'' and what exactly
is ``data''? While some definitions have been proposed, many of the laws
throughout the world are based on a separation between content data and
non-content data.\footnote{``Electronic evidence is any data resulting from the
  output of an analogue device and/or a digital device of potential probative
  value that are generated by, processed by, stored on or transmitted by any
  electronic device'' \citep[p.~561]{Biasiotti18}.} The latter is known also as
meta-data. It has usually received less protections from law~\citep{Daskal16},
although it has long been important for both surveillance and evidence
collection. Traditionally meta-data covers things like Internet Protocol (IP)
addresses and domain names, but \change{today} it extends also to location data
based on the Global Positioning System (GPS) or the location of cell towers. In
addition, many laws further talk about traffic data (non-content) and subscriber
data (content), the latter including things like electronic mail addresses,
credit call details, and phone numbers of consumers who have subscribed to a
particular service. But due to changes in telecommunications and information
technology in general, such as the rise of cloud computing, caching,
load-balancing, and content delivery networks, it is difficult to assess when
data is in motion (cf. meta-data and traffic data) and when it is at rest
(cf. content data). To this end, there have been frequent calls to better
separate different data types in jurisdictions~\citep{Biasiotti18, Svantesson16,
  Warken20}. European data protection laws have further complicated the
situation with the introduction of the notion of personal data. These issues
with different notions of data motivate the present work to take a brief look on
how technology companies themselves classify their data when releasing it to
LEAs. The transparency reports provide the empirical material for the brief
examination. In addition to the data types, a few points are also made about the
legal practices of the companies.

The remainder of the paper has a straightforward structure. The opening
Section~\ref{sec: background} discusses the legal and practical background
behind cross-border access to data by law enforcement agencies. The subsequent
Section~\ref{sec: challenges} outlines the contemporary controversies,
challenges, and responses to these. Both sections are based on a review of
existing literature; the goal is not to engage in detailed legal analysis but to
provide an overview of the recent trends and the research of these. In other
words, the focus is on the policy side of things. Then, Section~\ref{sec:
  insights} presents the noted brief empirical analysis of the transparency
reports. The conclusion and a discussion follow in the final Section~\ref{sec:
  conclusion}.

\section{\change{The Tradition of Cross-Border Data Access}}\label{sec: background}

Law enforcement agencies can access online data in many ways \change{for their
  crime investigations and foresight}. Much of online data is openly available
in the Internet. People throughout the world share their personal data via
social media and other technology platforms. Hence, LEAs and other state
agencies have long used so-called open source intelligence to gather big data
insights about various issues, \change{such as public opinion, harmful online
  content, hate speech, and disinformation. If anything, such open source
  intelligence has become more important in recent years. Though, at the same
  time, the} vast amounts of publicly available personal data have raised the
usual concerns about surveillance, privacy, and the legal qualifications of
public data as electronic evidence~\citep{Lewulis22, Scassa17, Trottier15,
  Warken20}. \change{However, the situation is different} with closed online
data held by private sector companies.

\change{Traditionally}, law enforcement agencies \change{in the US} have
\change{had} four options to access closed online data: different \textit{ad
  hoc} arrangements enacted directly with companies, letters rogatory, law
enforcement cooperation, and mutual legal
assistance~\citep{Westmoreland15}. \change{Although the situation has been
  different in Europe, the four options are still important for elaborating the
  historical background of cross-border data access practices. In particular, it
  is commonly presumed in the literature that the first option has been the most
  commonly used in recent years.}

With the first option, LEAs request closed online data from technology companies
or their subsidiaries directly.\footnote{~It should be further mentioned that
  LEAs are not the only ones making these direct requests; also private
  companies, associations, and even individuals (typically involved in civil or
  criminal proceedings) are occasionally making inquiries about disclosing
  non-public (personal) data. Copyright infringements are a good example in this
  regard.} Since most of the large technology companies are located in the
\change{US}, this option has become almost a standard procedure for European and
other law enforcement agencies to access closed online data \citep{Abraha21,
  DeHert18}. These direct requests are also known as unilateral actions
\citep{Abraha21} or unmediated access practices~\citep{Carrera15}. Such
practices align with the increased sharing of data between private companies and
public sector bodies particularly in the domains of cyber security and cyber
crime through the establishment of different, often informal public-private
partnerships~\citep{Bossong17, Ruohonen20EJSR}, which have also raised concerns
about legal accountability, data protection rights, and other related issues
particularly regarding the private sector partners~\citep{Christou18,
  Shillito19}. Private sector technology companies have also long been
pro-active in tackling cyber crime on their own, without the involvement of
LEAs; an approach that has led to some controversies over whether the companies
have overstepped their roles~\citep{Holt18}. \change{There are also other issues
  with the direct request practice.}

\change{In particular, it is frequently claimed in the literature that} the
direct requests by LEAs have put technology companies to a role in which they
alone evaluate privacy and human rights concerns; a task many companies are
ill-prepared to face already due to a lack of expertise and
resources~\citep{Daskal16, DeHert19}. \change{Although it is difficult to
  evaluate how well this claim generalizes internationally, the concern can be
  assumed to be valid particularly regarding countries with poor human rights
  records and weak rule of law provisions. There is also a corollary to this
  claim. When} companies are trying to protect their users against undue
interference by public authorities, certain parts of the criminal justice
systems have effectively been
privatized~\citep[\change{cf.}][]{Rojszczak22}. \change{The recent legislative
  changes and initiatives have indeed raised a concern about a potential
  privatization of criminal justice systems in Europe~\citep{Tropina22}.} From a
law enforcement perspective, on the other hand, the direct requests have
consequently led to a haphazard practice in which technology companies provide
different subsets of data in seemingly similar situations, making the overall
process unpredictable \citep{Svantesson16}. \change{Although almost all European
  countries have signed a legally binding international treaty on cyber crime
  that provides a legal basis for the direct requests, it is important to
  further remark that many countries in Europe still lack specific} national
laws specifying the legal provisions for the direct requests~\citep{EUROPOL20},
including the cases in which a direct request is made to a service provider
located in the territory of a given member state~\citep{Blazic20}. In other
words, there is neither proper judicial oversight nor a framework for protecting
the rights of people affected~\citep{CCBE19, Tosza20}. What is more, the direct
requests are also subject to domestic privacy and telecommunications laws. There
is always a risk of violating the law of a foreign jurisdiction when making a
direct request to disclose data.

Letters rogatory are used by a court in a foreign country to formally request
judicial assistance from another country. These are typically used through
diplomatic channels in the absence of assistance treaties or executive
agreements. In general, such requests have been inefficient and have had only a
limited role in law enforcement access to closed online data \citep{Efrat18,
  Kendall14, Westmoreland15}. Instead, much of the cross-border data access
occurs through the direct requests, law enforcement cooperation, and mutual
assistance treaties.

Law enforcement cooperation is also a common way for LEAs to access closed
online data. With this practice, agencies in one country share data they have
obtained through domestic processes with their foreign counterparts. Law
enforcement cooperation arrangements are typically based either on bilateral
relationships between countries or multilateral institutions such as
\text{INTERPOL}.\footnote{~That is, the International Criminal Police
  Organization.} There are also regional institutions such as EUROPOL and EC3 in
Europe, AMERIPOL in South America, and ASEANAPOL in Southeast
Asia.\footnote{~These are: the European Union Agency for Law Enforcement
  Cooperation, the European Cybercrime Centre, the Police Community of the
  Americas, and the Association of Southeast Asian Nations Police, respectively;
  see \citet{Guarda15} for organizations and initiatives on cyber crime in other
  regions. Also EUROJUST, enacted in 2002 for European co-operation in criminal
  matters, deserves a mention; see \citet{Bossong20} for other European security
  cooperation bodies.} The cooperation carried out through such institutions is
not strictly limited to law enforcement; intelligence agencies, customs,
anti-fraud organizations, and even coast guards may be involved. Multilateral
treaties have been also established through the United Nations, starting from
the 1988 Vienna drugs convention and including the 2000 Palermo convention
against transnational organized crime. Negotiations are also underway in the
United Nations for a new convention on cyber crime. There are also other
international treaties on cyber crime, such as the Budapest Convention, which
was signed in 2001 in order to facilitate cooperation between countries,
harmonize national criminal laws, and improve evidence collection and
investigation techniques. \change{As a legally binding multilateral treaty, the
  Budapest Convention also provides the legal basis for the direct request
  practice. In addition, numerous other related cooperation arrangements have
  been enacted}, such as the controversial Terrorist Finance Tracking Program
(TFTP) and its European counterpart.\footnote{~See \citet{DeGoede17} for a
  discussion about this program and its controversies.}

Finally, the mutual legal assistance treaty (MLAT) system has provided the
traditional way for LEAs to access closed online data without infringing
countries' territorial sovereignties. The MLAT system is based particularly on
bilateral agreements through which one country requests the assistance of
another country within which the data and evidence are located. The system has
been popular; over eight hundred MLATs have been signed over the years
throughout the world~\citep{Efrat18}. The system generally only covers criminal
matters and law enforcement investigations; requests and matters related to
national security and intelligence are excluded, as are other domains such as
competition law enforcement, administrative proceedings, and civil
cases~\citep{Currie20, Martyniszyn15, Swire17}. Even with these exclusions,
there have been several MLATs between the \change{EU} and the \change{US} since
the 2001 terrorist attacks~\citep{Carrera15, Christakis21}. These range from law
enforcement cooperation on serious international crime and terrorism, the TFTP,
and exchange of personal data to extradition and exchange of passenger name
records of air carriers.  The same applies to other countries and allies of the
\change{US}, including countries such as India, which has made hundreds of MLATs
on cyber security and cyber crime with the \change{US} and other
countries~\citep{Ebert20}. \change{These points notwithstanding, the MLAT system
  has been under an increasing pressure in recent years.}

The MLAT system, which was designed and implemented long before the Internet,
has frequently been seen as being cumbersome, bureaucratic, and
inefficient---state agencies such as the Department of Justice in the
\change{US} may receive thousands and thousands of requests from foreign
countries, leading to a situation in which a request may take months or even
years~\citep{Abraha21, DeHert18, Maillart19, Shurson20,
  Swire17}.\footnote{~According to critical viewpoints, however, it is difficult
  to evaluate the empirical backing for these arguments since robust publicly
  available data is generally lacking about the MLAT system and the requests
  made via it~\citep{Carrera15, Siry19}. Many of the claims about MLATs raised
  in the literature \citep[e.g.,][]{Brown20, Christou18} are based on anecdotal
  evidence. See also \citet{Abraha19} for counterarguments to other common
  critiques of MLATs.} The explanation for such delays is simple: throughout the
world, crime today often involves electronic evidence and much of the evidence
is located in the \change{US}. Therefore, a crime involving online data
conducted in the soil of a European country may require a mutual legal
assistance request to the \change{US} within which a justification is also
required for the probable cause, which may not be the case in the \change{given}
European country. Due to the delays caused by the MLAT system, \change{it is
  often claimed that} the electronic evidence sought may have already
disappeared in the meanwhile \citep{Rowe20}, given that technology companies do
not typically retain non-content data for long periods of time due to economic
and other reasons~\citep{Karagiannis23}. Furthermore, the admissibility of
mutual legal assistance requests is usually subject to the dual criminality
principle, but since criminal laws and particularly cyber crime laws
\change{still} vary greatly across the world, refusals of the requests have
\change{allegedly} been common~\citep{Maillart19}.\footnote{~\change{Though,
    again, no publicly available empirical data exists for evaluating such
    arguments.}} These examples help to understand why European and other LEAs
have often opted for the direct requests.

\section{Challenges, Controversies, and Responses}\label{sec: challenges}

There are many challenges, legal obstacles, and controversies about the access
to closed online data by LEAs. Cloud computing is among these. Most large
technology companies have their own clouds to which their customers and users
store personal and other data. Thus, also much of the evidence sought by LEAs is
in these clouds. The fundamental problem is that the clouds involve multiple
jurisdictions. This problem affects data protection~\citep{Celeste21}, criminal
justice, and technology companies themselves who must be familiar with multiple
national laws. Oftentimes, moreover, cloud service providers scatter their data
and data replicas across multiple servers located in multiple countries in order
to optimize performance and guarantee reliability and
availability~\citep{Shakrami21}. In such cases a cloud service provider itself
may not know where data is stored at any given moment.

To make some sense of the complex judicial questions involved, many cloud
service providers generally only provide data directly requested by LEAs in case
these perceive a given LEA as competent, its claims justified, and the request
having a legal basis and a valid purpose~\citep{Balboni13,
  Kahvedzic16}. However, the practices vary between companies; some companies
may even notify their users that their data was requested by a LEA. If a request
is denied for voluntary disclosure of data, a cloud service provider may require
a legal process involving a warrant, a court order, or a
subpoena~\citep{Svantesson16}. After the Snowden revelations, such requirements
have become common for companies located in the \change{US}; the practice of
quietly complying with requests has supposedly ended~\citep{Kahvedzic16}. For
these and other reasons, as already noted, the practice of direct requests is
generally incoherent and problematic for European LEAs. But the MLAT system is
not perfect either: in the context of cloud computing and related technologies,
a LEA may not even know to which country a mutual legal assistance request
should be directed~\citep{Abraha20, Maillart19}. Cloud computing involves also
many other issues.

Among other things, cloud computing has brought a further challenge: following
technical standards, guidelines, and good security practices, many companies
encrypt personal and non-personal data they store to their clouds. Sometimes
data is encrypted in such a manner that the companies themselves do not have
retrieval possibilities. Due to concerns about the capabilities of LEAs and
intelligence agencies for decryption of many common protocols, end-to-end
encryption has furthermore gained widespread traction in recent years. The
challenges for law enforcement agencies are well-known~\citep{Casino22,
  Ryder16}. To tackle the challenges, LEAs and intelligence agencies in some
countries have turned toward so-called offensive security and government
hacking; the use of exploits to compromise services and devices, the remote
installation of malware, and other opaque, less accountable, and controversial
means of accessing data and tackling cyber crime~\citep{BellovinLandau16,
  Brown20, Daskal16, Koops18, Ruohonen19JITP}. Some other countries, including
Australia in particular, have introduced specific but controversial
anti-encryption laws~\citep{McGarrity20}. Furthermore, data localization laws
have been enacted in some countries.\footnote{~See \citet{Abraha21},
  \citet{Celeste21}, and \citet{Selby17} for these data localization laws and
  their controversies.}  Yet the general challenges brought by cloud computing
are not limited to encryption, government hacking, and data localization.

\change{With respect to} jurisprudence, cloud computing has brought the
difficult questions about state sovereignty, territoriality, and
extraterritoriality to the forefront of debates. \change{The legal arguments
  raised in the debates often involve international law} and states' explicit
consent for other states to access their sovereign, territorial
realms. \change{Outside of jurisprudence}, the international debates have been
either explicitly or implicitly about reforming the MLAT system. Some progress
has also occurred.


Notably, the \change{US} passed the Clarifying Lawful Overseas Use of Data Act
(CLOUD Act) in 2018.\footnote{~\change{H.R.4943}.} It amends the older Stored
Communications Act (SCA), \change{which is a part of the Electronic
  Communications Privacy Act (ECPA) of 1986.}\footnote{~\change{18 U.S.C. § 2701
    to 2712.}} The CLOUD Act was a response to a famous case involving
Microsoft; the company challenged a warrant issued for turning emails stored to
a cloud service located in Ireland, arguing that the SCA did not allow companies
in the \change{US} to disclose data stored to foreign locations.\footnote{~See
  \citet{Abed19}, \citet{Carrera15}, \citet{Daskal18}, \citet{Shurson20}, and
  \citet{Siry19} for discussions about this famous and well-analyzed Microsoft
  case.} The CLOUD Act, which was passed before the \change{US} Supreme Court
had a second say on the case, resolved the question about extraterritoriality;
now federal LEAs in the \change{US} have a right to compel companies based in
the \change{US} to provide requested data stored to cloud computing servers,
irrespective of their territorial location.\footnote{~See \citet{Abraha20} for a
  detailed analysis of the CLOUD Act.}

In terms of international implications, the CLOUD Act is notable also because it
further encourages foreign governments to enter into executive bilateral
agreements with the \change{US}, allowing a qualified foreign government to
access any type of data (including content data) held by companies located in
the \change{US} without going through the cumbersome MLAT
system~\citep{Abraha21, Daskal18b}. Thus far, however, not many countries have
made such agreements; the United Kingdom (\change{UK}) is the notable exception,
although also Australia is about to strike an agreement. \change{It is likely
  that also other countries in the so-called Five Eyes alliance will sign
  similar agreements. There is also} an important incentive for making such
agreements: the SCA and the \change{ECPA} forbid companies in the \change{US}
from disclosing content data to foreign countries without a warrant issued by a
court in the \change{US} via a MLAT, although non-content data can be provided
on voluntary basis~\citep{DeHert18, Shurson20, Tosza20}. This restriction has
been seen as problematic for European LEAs and their
investigations~\citep{Christakis21}, although, according to recent surveys,
basic subscriber information and traffic data are much more often sought than
content data \citep{EUROPOL20}. In any case, as briefly discussed later on, the
\change{US} has encouraged also European countries to make bilateral agreements
based on the CLOUD Act.

Somewhat similar MLAT-related reforms have been pursued also in
\change{Europe. Interestingly, there has also been a visible policy-making
  cleavage in the European efforts to draft new criminal laws: the Council of
  Europe has actively and successfully promoted the Budapest Convention
  globally, while, at the same time, reforms have been pursued also in the
  EU.}\footnote{~\change{It is worth emphasizing that} the initial steps in Europe
  go as far as the 1950s. Notable milestones include the European Convention on
  Mutual Assistance in Criminal Matters of the Council of Europe of 20 April
  1959 and its 2000 supplement, the Mutual Assistance in Criminal Matters
  between the Member States of the European Union. Judicial cooperation in
  criminal matters was also specified in Article 82 of the Treaty on the
  Functioning of the European Union (TFEU). This article in the TFEU has
  provided the legal basis for the recent reform proposals for electronic
  evidence.} \change{Regarding the latter}, in order to harmonize national
criminal laws across the \change{EU}, the European Parliament and the Council
later adopted Directive 2014/41/EU regarding the European Investigation Order in
criminal matters.\footnote{~See \citet{Blazic20}, \citet{Knytel20},
  \citet{Tosza20}, and \citet{Warken20} for analyses of this directive and its
  national adaptations.}  Based on a notion of mutual recognition, this
directive replaces the previous European MLAT system by introducing a unified
legal instrument covering all types of cross-border investigative
measures. Among other things, this instrument covers cross-border interception
of the content of telecommunications as well as access to bank accounts and
financial transactions after an agreement by an executing member
state~\citep{Braum21}. However, it does not lay down provisions for electronic
evidence collection. There also remains many issues with the directive's
national transpositions, and, moreover, \change{Denmark and Ireland} have opted
out from its implementation altogether.

To this and other ends, the European Commission proposed in 2018 a new
``e-evidence regulation'' and an associated directive for cross-border
investigations in the union.\footnote{~COM/2018/225 final and COM/2018/226
  final.}  Based on arguments that it takes too long to gain electronic evidence
across borders, it specifies judicial requests that can be served directly to
service providers located or represented in the European
Union.\footnote{~SWD/2018/119 final.} These service providers include Internet
service providers, social networks, online marketplaces, domain name registries
and registrants, so-called information society services such as cloud hosting
services, and others. In essence, the providers of these services must then
disclose any data deemed as evidence, regardless of the territorial location of
the data~\citep{Christakis21, Karagiannis23, Warken20}. All essential data types
are covered; subscriber data, access data, transactional data, and content
data. As has been pointed out~\citep{EDPS19, Rojszczak22}, these categorizations
conflict with the concept of traffic data specified in the e-privacy Directive
2002/58/EC as well as with the new categories considered in the proposed drafts
for a new amending e-privacy directive.

Due to the proposal's extraterritorial reach, European LEAs would also have
access to data of citizens outside of the European Union; an implication that
may cause political problems and conflicts with other
jurisdictions~\citep{Tosza20}. Regarding the United States, particularly the
SCA's and ECPA's blocking provisions on content data may be problematic for
implementing the planned e-evidence regulation~\citep{Casino22,
  Shurson20}. Furthermore, there are data protection and many other issues
involved. When compared to the earlier Directive~2014/41/EU, the proposal has
much weaker provisions for fundamental rights; the overall provisions for
extraterritorial data access derive from the viewpoints of law enforcement
agencies~\citep{Bose18, CCBE19, DeHert18, Karagiannis23, Tosza20}. As explicitly
noted in the proposal, the issues extend well-beyond data protection and
privacy; also other fundamental rights are affected, including the right to
freedom of expression and assembly, the right to defense, the right to an
effective remedy and a fair trial, the right to freedom to conduct a business,
the right to liberty and security, and so forth. According to the proposal, it
is primarily the service providers who must evaluate whether an order violates
these rights~\citep{Rojszczak22}. Thus, the proposal has the same issues as the
direct request practices; a provider should be familiar with the nature of data
it possesses and releases, it should know data protection and human rights laws
as well as the domestic criminal laws of the requesting country, and so
forth. The proposal further imposes strict time limits for providers to comply
and subjects them to sanctions in case of non-compliance. The legal grounds on
which a provider may oppose an order are also weak. In general, serious concerns
have thus been expressed about how well service providers are able to fulfill
their protective function~\citep{Bose18}. Given the lack of harmonization across
the member states on criminal offenses, \change{a concern has been also raised
  about} the dual criminality principle, which is an essential safeguard for the
fundamental rights~\citep{EDPS19}. \change{In theory}, a member state
\change{could} rely on the assistance of another member state to enforce
criminal sanctions that do not exist in the criminal law of another member
state.

It also remains unclear whether European citizens would be allowed to challenge
the evidence collection decisions through the Court of Justice of the European
Union (CJEU) or national courts should they believe that their rights have been
violated~\citep{Rojszczak22, Siry19}. As with some other related European
laws~\citep{Jesserand18}, there is also a question whether citizens should be
notified about access to their data by LEAs during or after their
investigations; currently, the notification mandates vary across the member
states~\citep{Kusak17}. All in all, the implications of the e-evidence proposal
are thus comparable to those of the CLOUD Act; the fundamental rights are
potentially violated, and European LEAs should have access to both European and
foreign cloud services according to this proposal, which, however, is still
under policy-making in the European Union.

As is well-known, there has been a long-standing tension between the United
States and the European Union on matters related to cross-border data access,
surveillance, privacy, and data protection. The notable events in Europe include
the inclusion of data protection to the Charter of Fundamental Rights of the
European Union (CFR) in 2009, the enactment of the General Data Protection
Regulation (GDPR) and the so-called ``police'' Directive (EU) 2016/680 in 2016,
the decision of the CJEU to annul the so-called Safe Harbor agreement in 2015,
and the subsequent decision of the court to further invalidate the so-called
Privacy Shield framework in 2020. All are important events regarding
transatlantic data flows and the associated surveillance. Given this
background---and unlike in the United States~\citep{ACLU18, EFF18}, it is
surprising that the CLOUD Act and its extraterritorial reach created only a
little public \change{concerns expressed by} European policy circles and civil
society groups~\citep{Obendiek21}. The same point applies to the e-evidence
proposal; the voice from European civil societies \change{was} largely absent
during the policy-making of the planned regulation \change{(though, see
  \citealt{EDRi23})}. On the other hand, the position of \change{the CJEU} has
been consistent; the legal frameworks and practices in the United States fail to
protect the personal data of Europeans. Beyond saying that, it is beyond the
scope of this paper to delve into this pressing transatlantic
topic.\footnote{~See \citet{Swire17} or \citet{Setty15} on the surveillance laws
  in the United States and \citet{Celeste21}, \citet[][pp.~56--63]{Naef23}, or
  \citet{Murphy21} for analyses of the two so-called Schrems decisions on
  transatlantic data flows. There is also a new transatlantic data pact solution
  in the making, but it is too early to speculate about its legal, political,
  and other implications; see \citet{Manancourt22} for the recent political
  events.}  Instead, a few final points are warranted about cross-border data
access and the MLAT system.

Thus, to briefly continue, there have been several recent talks between the
United States and the European Union on cross-border access to electronic
evidence by LEAs. In line with the long tradition of promoting bilateral
MLATs~\citep{Harfield03}, the United States has tried to push the CLOUD Act as a
solution for cross-border data access, suggesting that it can be either used for
signing bilateral agreements between European countries and the United States or
tied directly to the planned e-evidence regulation in the European
Union~\citep{Daskal18a}. Such suggestions have received some skepticism on the
European side.

While the European Commission has been eager to present its e-evidence proposal
as a solution that would solve the conflicts in law between different
jurisdictions, suggesting an executive agreement with the United States and its
CLOUD Act, the European Parliament's LIBE committee has raised a number of
concerns.\footnote{~Regarding the executive agreement see COM/2019/70 final.}
Among these is a concern about reciprocity: the CLOUD Act places restrictions
for a qualified foreign government from requesting data about the citizens of
the United States, whereas, at the same time, the United States can request all
data irrespective of territories~\citep{EP19}.\footnote{~The same point has been
  raised also regarding other countries such as India~\citep{Annappa22}.}  There
is also a question whether the European Union has the competency (as opposed to
the individual member states) to strike a single executive agreement with the
United States. Analogously, it is unclear whether the CLOUD Act's notion of a
qualified government can apply to the European Union as a
whole~\citep{Abraha21}. Furthermore, the chief data protection authorities in
the European Union have raised concerns that the CLOUD Act does not provide
enough guarantees for the protection of personal data and the fundamental rights
thereto, which, according to their position, should be taken into account when
reforming the MLAT system through a unified European solution (see
\citealt{EDPB19a} and, analogously, also \citealt{CCBE19}). There are also
potential conflicts between the CLOUD Act and the GDPR, particularly with
respect to the latter's Articles 48 and 49 on transfers or disclosures not
authorized by the European Union law and derogations for specific
situations~\citep{Christakis19, Daskal18, Siry19}. Given the ongoing issue with
transatlantic data flows, also the GDPR's Articles 45 and 46 have been seen as
potential problem sources~\citep{Casino22, Shurson20}. When the European
Parliament adopted the Commission's proposal for the e-evidence regulation in
2020, it also clarified issues with the GDPR's purpose limitation specified in
Article~5 and proposed that data subjects should be informed and allowed the
right to seek a remedy~\citep{EP20}. But, at the time of writing, much remains
open about the proposal's relation to the CLOUD Act. Several scenarios and
potential solutions are possible~\citep{Christakis21, Rojszczak22}. As the
international implications remain unclear, it suffices to continue by providing
a few empirical insights about the practices of technology companies.

\section{Empirical Insights}\label{sec: insights}

A few quantitative points can be made by using a small dataset assembled from
the transparency reports released by technology companies. The transparency
index maintained by the non-governmental \citet{AccessNow23a} organization was
used to select the companies. In total, the dataset covers transparency
information collected from 71 companies. These are shown in Appendix; see
Tables~\ref{tab: sample 1} and \ref{tab: sample 2}. Most of the companies have
their headquarters in the United States, but there are also a few companies
located in Europe, Canada, New Zealand, and other countries. Most of the
companies can be further classified to a category of companies providing
Internet and mobile ecosystems. In addition, there are 15 telecommunications
companies, and two special cases. The transparency reports were released mostly
during the past three years or so.

\begin{figure}[p!]
\centering
\includegraphics[width=\linewidth, height=18cm]{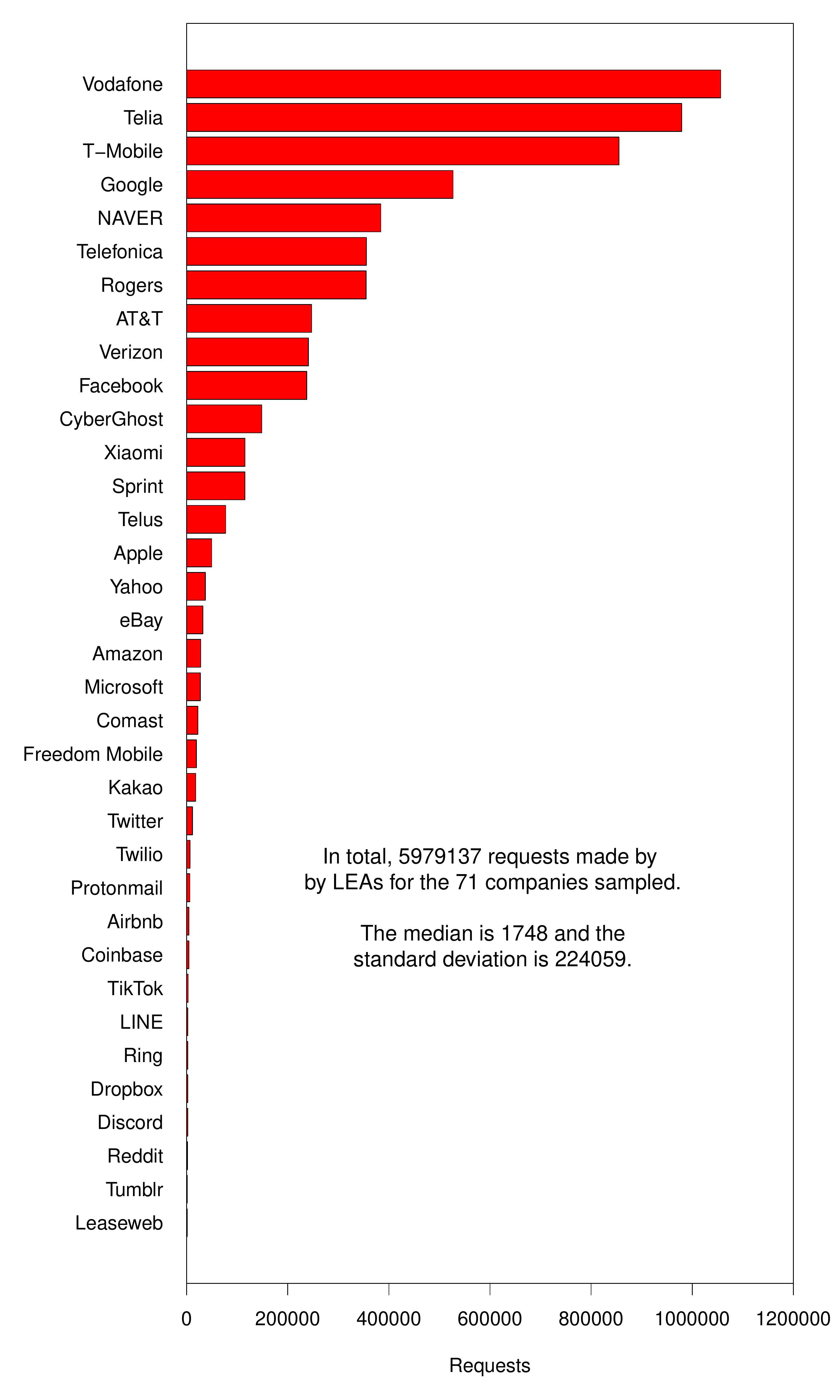}
\caption{Data Access Requests for the 35 Most Frequently Requested Companies}
\label{fig: requests}
\end{figure}

The amount of requests made by LEAs for the closed online data held by the
companies sampled provide a good way to start the quantitative exposition. Of
the companies sampled, 64 provided quantitative data on the requests. Thus, the
raw request counts are shown in Fig.~\ref{fig: requests} for the 35 most
frequently requested companies. These cover both domestic and international
requests as well as requests made to the companies' subsidiaries located in
different countries. As noted by many companies, LEAs make also so-called
emergency requests and preservation orders to freeze data for specified amounts
of time. These are covered in the counts. On the other hand, content blocking
requests, copyright takedown queries, and related requests are
excluded. Regarding the United States and a few other countries, also requests
made by intelligence agencies are included.\footnote{~In the United States these
  requests cover so-called national security letters and requests made based on
  the Foreign Intelligence Surveillance Act (FISA) of 1977.} As noted by some
companies, state agencies in the United States have instructed the companies to
report these requests by using ranges; the upper limits were used for these
cases. However, not all companies report these surveillance requests. Many
companies located outside of the United States also note that local laws prevent
them from reporting such national security and intelligence requests.

Given these clarifications, it can be concluded that data from
telecommunications companies is heavily requested by LEAs. Vodafone, Telia, and
T-Mobile have been the top-three most frequently requested companies. Google
comes only as the fourth and Facebook as the tenth in the ranking shown in the
figure. Microsoft does not even make it to the top-15. Also other
telecommunications companies, such as Telef\'onica, Rogers, AT\&T, and Verizon,
are frequently requested to release data. Given that many of these
telecommunications companies are based in Europe, these observations seem to
contradict some recent surveys according to which Google, Facebook, and
Microsoft would have been the most frequently contacted companies by European
LEAs in their investigations~\citep{EUROPOL20}. Another point is that the large
amount of requests made to telecommunications companies casts some doubts over
whether the existing literature might have slightly exaggerated the problems
associated with cloud computing. In any case, three further points can be made
about the practices of the companies for releasing data:
\begin{itemize}
\itemsep 5pt
\item{About 69\% of the companies sampled explicitly state that they require a
  proper judicial process before releasing their data; the remaining ones do not
  mention anything about legal procedures. While the exact procedures vary from
  a jurisdiction to another, typically warrants, court orders, or subpoenas are
  mandated by the companies. However, many of the companies that require a
  judicial process mention that they still release some data also on voluntary
  basis should they perceive a request as valid. Furthermore, only a few
  companies provide some vague details regarding the cases where they have
  refused to comply with requests to disclose data.}
\item{About 42\% of the companies explicitly state that they have specific
  mandates for international requests; the remaining companies do not mention
  anything about these cross-border requests. The companies who have mandates
  typically point toward MLATs. Also letters rogatory are occasionally
  mentioned. Some companies further mention the bilateral agreements based on
  the CLOUD Act. That said, together with the previous point, the relatively
  small amount of explicit mandates for cross-border requests implicitly
  indicates that LEAs likely still use the problematic direct request
  practice. The general lack of rigor in the transparency reports allows to
  also question how well the companies understand existing laws about data
  protection, privacy, human rights, and criminal justice in foreign countries.}
\item{About 48\% of the companies mention that they try to notify their users
  about requests made by LEAs; the remaining ones do not mention anything
  regarding such notifications. Of the companies who seek to notify their
  customers and users, many mention that they abstain from notifications in case
  a court-backed gag order is placed, a notification involves an emergency
  situation, such as a danger of death or a serious physical harm to a person or
  place, or in case of other sensitive issues such as sales of lethal drugs.}
\end{itemize}

The transparency reports provide also some interesting insights about specific
data types and categorizations of data. The companies can be categorized into
five groups in terms of these data types and categorizations. The first is a
problematic group in terms of transparency: about 31\% of the companies do not
specify any details about what kind of data they possess and disclose upon
requests made by LEAs.

The second group, which consists about 20\% of the companies sampled, specifies
releasing content data based on user accounts. Such content varies from a
company to another. Some companies in this group maintain that they do not
retain any non-content data due to their so-called no-logs policy. There is also
an interesting outlier in this group: a company called 23andMe, which provides
deoxyribonucleic acid (DNA) tests for consumers. The content from such tests
clearly falls to the GDPR's category of sensitive data, and, for whatever
reason, LEAs in the United States have recently made eleven requests to the
company for releasing such genetic data.

The third group accounts for about 24\% of the companies sampled. This group
relies on the traditional distinction between content data and non-content
data. There are some notable companies in this group, including Amazon, Cisco,
Cloudflare, Dropbox, Facebook, and IBM, to name a few examples.  As with the
second group, both types of data vary from a company to another. For instance,
Amazon classifies subscriber information (such as names, addresses, and billing
information) into a category of non-content data, while mainly reserving the
category of content data to all data stored and processed in the company's cloud
hosting service---to which LEAs in the United States have access with the CLOUD
Act irrespective of the territorial location of a particular data center. The
same applies to IBM; content data refers to processing, storage, and hosting of
data. There is also one telecommunications company in this group: Comcast, which
maintains that the content of communication (such as the body of an email or a
telephone conversation) are content data, whereas non-content data includes such
things as header information, list of phone numbers or email addresses, and
signaling, addressing, or routing information. Another good example is Zoom
whose content data includes the content of video teleconferencing calls, among
other things.

The fourth group contains three companies: Apple, Uber, and Xiaomi. With
slightly different terminology, these three companies separate three data types:
device and vehicle data, data based on financial identifiers, and content data
behind user accounts. Releasing device data includes requests made based on
serial numbers, IMEIs, MEIDs, and IMSIs.\footnote{~International Mobile
  Equipment Identity, Mobile Equipment Identifier, and International Mobile
  Subscriber Identity, respectively.} Uber is also an interesting case because
the company is very data-hungry. There is thus a treasure drove of data for
LEAs; both drivers' and customers' device data, vehicle information (including
license plate data and vehicle insurance data), billing information (including
credit card details and financial transactions), trip information (including
pick-up and drop-off locations as well as GPS data in some cases), records of
communication, customer service records, details about drivers (including their
photos and drive license copies), data about auxiliary functionalities
(including data about restaurants and orders placed in these), and freight data
(including calls logs between Uber and carriers, load information, and GPS
data). However, none of the three companies disclose details about what the
device data actually contains; in principle, this type of data may refer to
content of communications (such as the content of telephone calls or text
messages) as well as to content data stored directly into a device, whether
photos, files, or something else.

The final group contains rest of the telecommunications companies alongside a
few companies providing Internet and mobile ecosystems. All of these acknowledge
providing wiretaps to communications. In the United States these are known as
so-called ``pen registers'' and ``trap and trace devices'', which are also
provided by companies in the other groups, including Yahoo, Cloudflare, and
others. These allow capturing real-time information about non-content data, such
as the sources and destinations of communications, the duration and size of
communications, and so forth. Other telecommunications companies further allow
lawful real-time interception of the content of communications based on
applicable national laws. As for data types and categorizations of data, there
is some variance in this group. For instance, Telef\'onica separates content
data obtained via lawful interception from meta-data, further noting that LEAs
make also content blocking requests and requests to suspend services. The
Swedish Telia further separates lawful interception data from historical
archival data and subscription data. The historical data is worth emphasizing
because of the long list of controversies and legal cases involving the
corresponding data retention laws in the European Union.\footnote{~For instance,
  the CJEU invalidated the data retention Directive 2006/24/EC in 2014, but many
  member states have still kept their own national laws for retaining meta-data
  on telecommunications (\citealt{Fennelly19}; \citealt[pp.~44--51]{Naef23}). In
  general, extensive meta-data archives are still today maintained about
  Internet and mobile telecommunications throughout much of Europe.} Unlike the
other companies sampled, Telia is also explicit about its lack of power to infer
about data gathered by intelligence agencies with their signals intelligence.

\section{Conclusion}\label{sec: conclusion}

The literature reviewed agrees on one thing: the current situation is
problematic for various of reasons. The problems are well-recognized also by the
stakeholders and practitioners involved. Governments, law enforcement agencies,
technology companies, professional associations, data protection authorities,
civil society groups and privacy activists, and others have all expressed
concerns. 

The practice of directly requesting data from companies is particularly
problematic in the context of cross-border electronic evidence gathering. As was
noted, this practice is incoherent and non-optimal for law enforcement agencies
in their legitimate cross-border investigations because companies themselves
largely dictate what they are willing to voluntarily release under domestic
laws. The same applies from a perspective of fundamental rights: companies
should not be the ones who evaluate privacy, data protection, and human rights
issues. Furthermore, the direct request practice largely lacks proper judicial
oversight in Europe and elsewhere. Similar points apply to cross-border evidence
collection and data sharing through law enforcement cooperation, the e-evidence
proposal, and public-private partnerships. In particular, there are various data
protection and other issues with the Budapest Convention, which, among other
things, includes provisions for extraterritorial cross-border data access with
only a few restrictions.\footnote{~In essence, the Budapest Convention allows
  access to data regardless of territorial boundaries and without any
  interference of third-parties, whether private or
  governmental~\citep{Lewulis22, Karagiannis23}. The second protocol of the
  Convention that was agreed in 2022 seems to further facilitate the direct
  request practice, although still upholding the MLAT system for some
  requests~\citep{Abraha19, Rojszczak22}. It is also worth noting that the
  Convention's notion of data seems to cover personal data covered by data
  protection laws~\citep{Balboni13}. To this end, the legacy data protection
  working party in Europe clearly stated that data controllers do not have a
  lawful authority to voluntarily disclose personal data they
  process~\citep{WP2917}. This statement applies equally to the Budapest
  Convention and the direct request practices. Also \citet{EDPB19b} later
  stressed that data controllers can normally only disclose personal data upon a
  proper judicial authorization. See also \citet{Carrera15} and \citet{DeHert19}
  about policy-making criticism regarding venue-shopping through the Council of
  Europe.} Here, the human rights concerns are graver, given that also countries
such as Azerbaijan and Turkey are parties to the convention. Against this
backdrop, there is some merit in the arguments that the CLOUD Act improves human
rights because state agencies in the United States instead of technology
companies evaluate the overall arrangements for the disclosure of
data~\citep{Daskal18b}. But as not many countries have thus far signed bilateral
agreements based on the act, the international situation still remains more or
less open. The empirical insights presented also hint that the direct request
practice is still used by LEAs. Finally, this practice is problematic also for
technology companies who are in a difficult position regarding compliance to
different laws in different jurisdictions, some of which may entail serious
sanctions. To this end, fears have been expressed that companies will resort to
data localization or persistent challenging in courts as
remedies~\citep{Daskal18, Shurson20}. According to a commonly raised argument
particularly in the United States, such solutions would generally challenge the
open and global nature of the Internet, allowing also easier surveillance for
domestic agencies, including those in authoritarian regimes.

The reform initiatives in Europe are still under policy-making. The existing
research literature has identified various issues and potential conflicts in law
particularly with respect to the Commission's proposal for the e-evidence
regulation and the potential executive agreement with the United States and its
CLOUD Act. Among these are: the competency of the European Union to sign an
executive agreement with the United States, the qualification of the whole
European Union as a qualified government as per the CLOUD Act, the e-evidence
regulation versus European data protection laws, the e-evidence regulation
versus the CFR, the e-evidence regulation versus the e-privacy directive, the
e-evidence regulation's coverage of citizens outside of the European Union
versus the CLOUD Act's restrictions regarding the citizens of the United States,
the e-evidence regulation versus the SCA's and ECPA's blocking provisions on
content data, the CLOUD Act versus the GDPR, the e-evidence regulation versus
admissibility to the CJEU or other courts, the e-evidence regulation and the
CLOUD Act versus the CJEU's decisions on the transatlantic data flows, and
probably more. As a trialogue is already underway for the e-evidence
regulation, it seems that many of these issues will be present also in the
future law.

While doing good work on the legal aspects, many authors have been shy to reveal
their preferences, abstaining from making policy recommendations. Though, there
are some exceptions; some authors have expressed enthusiasm, arguing that the
e-evidence proposal solves technical issues with evidence collection and
improves harmonization across the member states without threatening the
fundamental rights \citep{Blazic20}, while some others have been much more
skeptical, arguing that the risks of increased surveillance are not worth the
potential gains~\citep{Siry19} and that the fundamental rights are
threatened~\citep{DeHert18, Karagiannis23}. Given the enduring tug-of-war with
sovereignty and territoriality of the member states vis-\`a-vis European
criminal law~\citep{Knytel20}, it can be also remarked that the Commission's
e-evidence proposal threatens the territorial sovereignty of the member
states~\citep{Bose18}. It may thus be possible that some member states try to
opt out from the proposal's implementation. Furthermore, it is worth remarking
an argument that the criminal justice efforts by the European Union may put
political pressure on the national criminal justice systems, exacerbating
existing tendencies toward crime control and away from due
process~\citep{Murphy15}. In other words, the crises of democracy and rule of
law in some member states extend toward the common criminal justice and law
enforcement cooperation efforts at the European level~\citep{Rojszczak22}. These
concerns apply also to the proposed e-evidence regulation. More generally, it
seems reasonable to agree with \citet{Pawlak17} in that there is still a need to
strengthen due process safeguards, accountability conditions for investigative
powers, data protection, and oversight mechanisms.

Regarding international issues, those who have expressed critical viewpoints,
including data protection authorities, have generally argued that the MLAT
system should be reformed instead. Although problems prevail, the adoption of
the Directive 2014/41/EU for cross-border investigations in the European Union
is a step toward this reformist direction. There have been also some other
related efforts, including the INTERPOL's e-MLA initiative and the e-CODEX
project in the European Union, both of which are generally designed to increase
the performance of the MLAT system through secure platforms for exchanging
requests, as well as the designations of specific personnel as national contact
points and liaison officers in different jurisdictions~\citep{Abraha21,
  Casino22, Renard18}. In general, those in favor of retaining and reforming the
MLAT system perceive the system---even with its many flaws---as a still better
option compared to countries' unilateral exercises of extraterritorial
jurisdiction over data stored in foreign territories.

Four further points can be raised. The first is about cloud computing. The CLOUD
Act and the e-evidence proposal both seem to solve many of the problems for law
enforcement agencies and their investigations. However, the empirical insights
presented allow to question how pressing the problems truly are; the clear
majority of requests by LEAs is targeted toward traditional telecommunications
companies. Wiretaps, historical meta-data archives, and other means of more
traditional evidence collection are still widely used. Furthermore, the new
legislations and proposals have not addressed the other problems with cloud
computing. Encryption remains an issue, data localization is still on the agenda
as are anti-encryption laws, and government hacking has likely only increased in
recent years.\footnote{~See COM/2022/500 final for government hacking.} The
emergence of end-to-end encryption widely used in mobile applications has
intensified the calls for these controversial remedies.\footnote{~See
  COM/2022/209 final for an anti-encryption proposal \change{in the EU. This
    European proposal aligns with similar ones elsewhere, including the
    so-called EARN IT Act in the US and the Online Safety Bill in the UK.}} There
is also a more philosophical or political viewpoint to cloud computing: the rise
of technology giants has brought a new ``digital corporate sovereignty'' against
which countries try to align their territorial jurisdictions and
sovereignties~\citep{Ruohonen21MIND}. The same viewpoint is shared by many LEAs
and criminal justice systems: they must regain their
sovereignty~\citep{Braum21}. These viewpoints have been common in the European
Union, which has recently been pursuing a new strategy under the label of
digital sovereignty.

The second point is about the data types and date categorizations. According to
the empirical insights presented, technology companies are not generally
consistent with their categorizations of data disclosed to LEAs. Although many
companies rely on the distinction between content and non-content data, there is
also a large amount of companies who seem to make no categorizations
whatsoever. What is more, personal data is commonly categorized both as content
data and non-content data. Also sensitive personal data is held and disclosed by
technology companies. In a sense, the old legal demarcations seem outdated not
only for electronic evidence collection but also for many data-intensive
companies. As the saying goes, data is merely data; with data mining it becomes
information, and with science information becomes knowledge. Thus, none of the
existing laws and proposals address such categories as behavioral data and
derived data~\citep[cf.][]{Schneier10}. Given the giant corporate empires that
have been built around data mining, it would seem reasonable to argue that
derived data about already well-profiled individuals would be valuable also for
LEAs when investigating, say, radicalization and violent extremism. This point
extends toward the increasing use of artificial intelligence by LEAs themselves,
which is another legal and ethical quagmire regarding data protection,
fundamental rights, accountability, and transparency.

The third point is about the transparency reports released by technology
companies. The analysis of these reveals many problems, from incoherence to a
lack of details and rigor. This observation supports earlier
results~\citep{Parsons19}. In general, it can be argued that accountability and
transparency requirements cannot and should not be outsourced to private sector
companies.

The fourth and final point is about extraterritoriality. Before continuing, it
should be emphasized that the unilateral assertion of extraterritorial
jurisdiction is nothing new as such. During the heydays of globalization, many
argued that the increased use of domestic criminal law extraterritorially was
unavoidable, necessary, and even desirable~\citep{Parrish12}. Similar arguments
have long been used in many civil cases involving extraterritorial
conduct. Domestic courts in various countries have been willing to exercise
their jurisdictions extraterritorially in as different domains as human rights,
financial regulation, intellectual property, and
anti-trust~\citep{Kahraman20}. Against this background, it is not surprising
that not only have the cross-border data access practices of LEAs involved
extraterritorial tenets, but all of the themes discussed in this paper have had
these same tenets.

The surveillance machinery that was revealed by Snowden is a prime example about
the exercise of extraterritorial power over sovereign
countries~\citep{Obendiek21}. Then, the Budapest Convention as well as the GDPR
both established clear extraterritorial provisions. In terms of the latter, the
regulation's Article~3 states that the protection mandate applies regardless
where the processing of personal data occurs. The same applies regarding the
CFR's Article~8 and the TFEU's Article~16. The CJEU has also extensively relied
on these extraterritorial provisions when enforcing European data protection
laws~\citep[pp.~55--92]{Naef23}. Extraterritorial arguments have been widely
used also by national courts in Europe and \change{the US}. Upon reviewing
these, \citet{DeHert18} identify twelve distinct arguments, including a strict
sovereignty argument (there is no legitimacy for cross-border data access
because a state's sovereignty is violated upon the unilateral exercise of power
by another state beyond its borders), an argument based on the physical location
of servers, the already noted MLAT-preference argument, an argument based on the
establishment of a business on a given territory, an argument based on the
nationality of data subjects, and so forth. In terms of domestic criminal laws,
these arguments correlate with the territoriality principle according to which a
perpetrator can be brought to a court in a country within which the crime was
committed, and the nationality principle whereby a court of a country has the
jurisdiction to prosecute a national of that country, irrespective of the place
where the crime was committed~\citep{Pocar04}. But regarding cross-border data
access by LEAs, both the CLOUD Act and the proposed e-evidence regulation seem
to rely on the argument about the location of businesses. For the former, it is
about companies based in the United States; for the latter, it is about the
establishment or representation of a business in any member state of the
\change{EU}.\footnote{~\change{Here, it is worth further remarking that also the
    \citeauthor{OECD22}'s \citeyearpar{OECD22} recent informal declaration notes
    that state agencies have a right under domestic laws to request private
    sector companies to disclose data that is not located within their
    territory.}} Both bypass the MLAT system, and both also continue the
extraterritorial tradition, allowing the United States and the European Union to
exercise their jurisdictions beyond their borders. While both seem to solve many
issues with cloud computing for law enforcement agencies in a relatively similar
manner, it is easy to agree with the arguments that an international solution is
needed in the long-run~\citep{Abraha21, Karagiannis23}. No country can alone
confront the pace of technological development and the crime that comes with it.

\section*{Acknowledgements}

This research was funded by the Strategic Research Council at the Academy of Finland (grant number 327391).

\section*{Statements and Declarations}

There is a conflict of interest with other researchers funded by the same grant (no. 327391) from the Strategic Research Council at the Academy of Finland.







\section*{Appendix}

\begin{table*}[th!b]
\centering
\caption{Sampled Companies and Their Transparency Reports (1/2)}
\label{tab: sample 1}
\begin{tabular}{lllc}
\toprule
Company & Country & Sector & Year \\
\hline
23andMe & The United States & Health (genetics) & 2023 \\
Adobe & The United States & Internet and mobile ecosystems & 2021 \\
Airbnb & The United States & Internet and mobile ecosystems & 2021 \\
Amazon & The United States & Internet and mobile ecosystems & 2022 \\
Apple & The United States & Internet and mobile ecosystems & 2021 \\
AT\&T & The United States & Telecommunications & 2022 \\
Cisco & The United States & Internet and mobile ecosystems & 2022 \\
Cludflare & The United States & Internet and mobile ecosystems & 2021 \\
Coinbase & The United States & Internet and mobile ecosystems & 2020 \\
Comcast & The United States & Telecommunications & 2022 \\
cPanel & The United States & Internet and mobile ecosystems & 2022 \\
CREDO & The United States & Telecommunications & 2020 \\
CyberGhost & Romania & Internet and mobile ecosystems & 2022 \\
Deutsche Telekom & Germany & Telecommunications & 2021 \\
Discord & The United States & Internet and mobile ecosystems & 2022 \\
Dropbox & The United States & Internet and mobile ecosystems & 2021 \\
eBay & The United States & Internet and mobile ecosystems & 2021 \\
Etsy & The United States & Internet and mobile ecosystems & 2021 \\
Evernote & The United States & Internet and mobile ecosystems & 2021 \\
Facebook & The United States & Internet and mobile ecosystems & 2022 \\
Freedom Mobile & Canada & Telecommunications & 2020 \\
GitHub (Microsoft) & The United States & Internet and mobile ecosystems & 2021 \\
Google & The United States & Internet and mobile ecosystems & 2021 \\
IBM & The United States & Internet and mobile ecosystems & 2021 \\
Kakao & South Korea & Internet and mobile ecosystems & 2022 \\
Kickstarter & The United States & Internet and mobile ecosystems & 2015 \\
Leaseweb & The Netherlands & Internet and mobile ecosystems & 2021 \\
Let's Encrypt & The United States & Internet and mobile ecosystems & 2022 \\
LINE & Japan & Internet and mobile ecosystems & 2022 \\
LinkedIn & The United States & Internet and mobile ecosystems & 2022 \\
Lookout & The United States & Internet and mobile ecosystems & 2020 \\
Lyft & The United States & Internet and mobile ecosystems & 2021 \\
MapBox & The United States & Internet and mobile ecosystems & Unknown \\
Medium & The United States & Internet and mobile ecosystems & 2015 \\
MEGA & New Zealand & Internet and mobile ecosystems & 2021 \\
Microsoft & The United States & Internet and mobile ecosystems & 2022 \\
NAVER & South Korea & Internet and mobile ecosystems & 2022 \\
Pinterest & The United States & Internet and mobile ecosystems & 2022 \\
Posteo & Germany & Internet and mobile ecosystems & 2022 \\
\bottomrule
\end{tabular}
\end{table*}

\begin{table*}[th!b]
\centering
\caption{Sampled Companies and Their Transparency Reports (2/2)}
\label{tab: sample 2}
\begin{tabular}{lllc}
\toprule
Company & Country & Sector & Year \\
\hline
Protonmail & Switzerland & Internet and mobile ecosystems & 2021 \\
Reddit & The United States & Internet and mobile ecosystems & 2021 \\
Ring & The United States & New technology & 2020 \\
Rogers & Canada & Telecommunications & 2021 \\
Slack & Canada & Internet and mobile ecosystems & 2021 \\
Snap & The United States & Internet and mobile ecosystems & 2022 \\
SpiderOak & The United States & Internet and mobile ecosystems & Unknown \\
Sprint & The United States & Telecommunications & 2019 \\
T-Mobile & The United States & Telecommunications & 2021 \\
TekSavvy & Canada & Telecommunications & 2022 \\
Telef\'onica & Spain & Telecommunications & 2021 \\
Telia & Sweden & Telecommunications & 2021 \\
Telstra & Australia & Telecommunications & Unknown \\
Telus & Canada & Telecommunications & 2021 \\
TikTok & China & Internet and mobile ecosystems & 2020 \\
Trade Me & New Zealand & Internet and mobile ecosystems & 2021 \\
Trustpilot & Denmark & Internet and mobile ecosystems & 2022 \\
Tumblr & The United States & Internet and mobile ecosystems & 2020 \\
TunnelBear & Canada & Internet and mobile ecosystems & 2020 \\
Twilio & The United States & Internet and mobile ecosystems & 2021 \\
Twitch & The United States & Internet and mobile ecosystems & 2020 \\
Twitter & The United States & Internet and mobile ecosystems & 2021 \\
Uber & The United States & Internet and mobile ecosystems & Unknown \\
Verizon & The United States & Telecommunications & 2021 \\
Yahoo & The United States & Internet and mobile ecosystems & 2021 \\
Vitru & The United States & Internet and mobile ecosystems & 2022 \\
Vodafone & The United Kingdom & Telecommunications & 2019 \\
Wickr & The United States & Internet and mobile ecosystems & 2022 \\
Wikimedia & The United States & Internet and mobile ecosystems & 2021 \\
Automattic & The United States & Internet and mobile ecosystems & 2021 \\
Xiaomi & China & Internet and mobile ecosystems & 2021 \\
Zoom & The United States & Internet and mobile ecosystems & 2022 \\
\bottomrule
\end{tabular}
\end{table*}

\FloatBarrier

\balance

\end{document}